\documentclass[a4paper]{jpconf}
\usepackage{graphicx}
\usepackage[mathscr]{eucal}
\usepackage[latin1]{inputenc}
\usepackage{amsmath}
\usepackage{amsfonts}
\usepackage{amssymb}
\usepackage{graphicx}

\usepackage{graphicx,amsmath,hyperref}
\usepackage{tikz}

\newcommand{\f}{\frac}

\def\tr{{\rm Tr}}

\newcommand{\bk}[2]{{\langle#1\,|\,#2\rangle}}

\newcommand{\bc}{\begin{center}}
\newcommand{\ec}{\end{center}}

\usepackage{wrapfig}
\begin{document}
\title{Spinfoam Cosmology\\ 
\hskip25mm
{\bfseries\itshape\selectfont quantum cosmology from the full theory}}

\author{Francesca Vidotto}

\address{Centre de Physique Th\'eorique de Luminy\footnote{Unit\'e mixte de recherche (UMR 6207) du CNRS et des Universit\'es de Provence (Aix-Marseille I), de la M\'editerran\'ee (Aix-Marseille II) et du Sud (Toulon-Var); laboratoire affili\'e \`a la FRUMAM (FR 2291).}
     , Case 907, F-13288 Marseille, EU;}
     
  \address{Dipartimento di Fisica Nucleare e Teorica,
        Universit\`a degli Studi di Pavia, and\\  Istituto Nazionale
        di Fisica Nucleare, Sezione di Pavia, via A. Bassi 6,
        I-27100 Pavia, EU.}

\ead{vidotto@cpt.univ-mrs.fr}

\begin{abstract}
Quantum cosmology is usually studied quantizing symmetry-reduced variables. Is it possible, instead, to define quantum cosmology starting from the full quantum gravity theory? In Loop Quantum Gravity (LQG), it is possible to cut the degrees of freedom in a suitable way, in order to define a cosmological model. Such a model provides a tool for describing general fluctuations of the quantum geometry at the bounce that replaces the initial singularity. I focus on its simplest version, a ``dipole" formed by two tetrahedra. This has been shown to describe a universe with anisotropic and inhomogenous degrees of freedom. Its dynamics can be given using the spinfoam formalism. I briefly review the present state of this approach
.
\end{abstract}

\section{Introduction}
I review here the main ideas of a new approach to quantum cosmology called \emph{spinfoam cosmology}. In order to do this, I need to start by recalling what is usually meant by \emph{quantum cosmology}. Where does the idea of quantum cosmology come from?


The original problem was the quantization of gravity. 
When the Wheeler-DeWitt equation was written down, it 
immediately appeared to be an intractable equation. 
Brice DeWitt and Charles Misner realized that one could study the quantization problem in a much simpler context: the finite dimensional dynamical system defined by cosmology.
In cosmology the symmeties assumed reduce the gravitational field to a system with one (or few) degrees of freedom (DOF). These are far easier to quantize than the full theory, of course. In this way we arrive to the Wheeler-DeWitt equation for cosmology, and to all the following minisuperspace models. The application of Loop ideas to this program, in the last decade, has lead to Loop Quantum Cosmology (LQC)\cite{Ashtekar:2008zu}, with its beautiful results (first of all singularity resolution \cite{Bojowald:2001xe,Singh:2009mz,Ashtekar:2009kx}) and therefore to a remarkable realization of the research program of DeWitt and Misner.

In the meanwhile, however, there have been main developments in the quantization of the full theory. Now the original problem of ``how to quantize gravity'' is rather well defined and LQG gives us several tool \cite{Rovelli:2010ly}. Therefore today we have the possibility to follow a different road to quantum cosmology: we can try to study quantum cosmology starting from the full quantum gravity theory. 

What can we obtain by exploring this road?
From a technical point of view, this could help us understanding how LQC is embedded in the full theory. But what I find more tantalizing is the possibility to have a tool to describe the quantum geometry at the bounce, namely in a regime where a single DOF could not be enough to capture the geometry. The other DOF of this model provide a natural way to include inhomogeneities \cite{Rovelli:2008ys,Battisti:2010kl}. One long term hope is to obtain a meaningful setting to describe the seeds of structure formation.


Here I describe a strategy to address these questions. 
LQC is imbedded in the full theory as an approximation. Therefore I discuss the approximations in the full theory. 
I review some basic features of the present model of LQG, describing at each step the approximation taken, and relating these approximations with cosmology.   Let me start from the approximation that is at the base of the birth of modern cosmology, as we can see it from this prospective.

\section{Embedding the cosmological principle in the graph Hilbert space}

Modern cosmology is based on the cosmological principle, that says that ``the dynamics of a homogeneous and isotropic space approximates well our universe''. The presence of inhomogeneities can be disregarded in a first order approximation, where we consider the dynamics as described on the scale of the scale factor, namely the size of the universe. Thus our approximation depends on the scale factor, it is not just a large scale approximation: 
it depends on the ratio between the scale factor and the interaction that we want to consider. If we consider the dynamics of the whole universe, this ratio gives 1, and an unique DOF is concerned. We can then recover the full theory adding these DOF one by one. We obtain an approximate dynamics of the universe, with a finite number of DOF.
%
%
%
%
%
\\[1mm]
{\bf Graph truncation}~~
In LQG there is a natural way to implement the cosmological expansion. This will become clear considering the kinematic.
In LQG the theory is defined on a direct sum of \emph{graph spaces} ${\cal H}_\Gamma=  L^2 [SU(2)^L/SU(2)^N]$, where the states are restricted to be the ones invariant under $SU(2)^N$ gauge transformations.
Each ${\cal H}_\Gamma$
 is an Hilbert space defined on a given graph $\Gamma$, and all the subgraphs and the graphs obtained by a permutation will be contained in such a space. 
 The graph captures the large scale DOF obtained averaging the metric
                  over the faces of a cellular decomposition formed by $N$ cells.
Working with graphs can be thought as taking an averaging of the metric on the faces of the cellular decomposition.
 These graphs can be characterized simply in terms of the number of nodes $N$, the number of links $L$ 
and their adjacency. 

A simple example of graph that can be chosen is the ``dipole'' 
$\Delta_2^*=\!\!\!\!\begin{picture}(20,8)
	\put(08,3) {\circle*{3}} 
	\put(30,3) {\circle*{3}}  
	\qbezier(8,3)(19,18)(30,3)
	\qbezier(8,3)(19,-3)(30,3)
	\qbezier(8,3)(19,9)(30,3)
	\qbezier(8,3)(19,-11)(30,3)
\end{picture}$~~~~\ 
with 2 nodes and 4 links. It is particularly nice because its dual $\Delta_2$ has a direct classical interpretation being a triangulation of a 3-sphere, formed by two tetrahedra glued together. This graph can be very useful in order to do calculation, so we will make use of it later, but our arguments can be generalized for all the graphs.

The truncation of the Hilbert space of LQG to the Hilbert space on a given single graph can be put in correspondence to the expansion in modes in cosmology, truncated at a finite order.
		This defines an approximated kinematics of the universe, 
		inhomogeneous but truncated at a finite number of cells. 
One can recover the full theory adding the DOF one by one. We start from the cosmological DOF, namely a single variable given by the volume, and then adding more DOF corresponds to add more nodes and links on the graph considered.
To obtain an homogeneous and isotropic space, one have to consider a regular graph.
%
%
%
%
%
\\[1mm]
{\bf Coherent states}~~~
On a graph space we can define an over-complete basis of  wave packets  \cite{Bianchi:2009ky}
$$    
\hspace{-2mm}    \psi_{H_\ell}(U_\ell) \!=\! \int_{_{SU(2)^N}}
\!\!\!\!\!\!\!\!\!\!
 dg_n ~
       \bigotimes_{\ell\in \Gamma} K
       (g_{n_{source(\ell)}}\!U_\ell\,
       g^{-1}_{n_{target(\ell)}}H_\ell ^{-1})
   ~~~~ \mbox{where}~~ g_n\!\in\!SU(2) ~~\forall\ \mbox{nodes}\ n
$$
and $\forall$ link $\ell$, the heat kernel $K$ peaks the holonomies $U_\ell$ on the $SL(2,\mathbb{C} )$ element $H_\ell$. This
\mbox{$
H_\ell\!=\! g_{\vec n}\ e^{-i ( \xi+i  \eta) \frac{\sigma_3}2}\ g^{-1}_{\vec {n}'}
$} determines two normals to the faces of the cellular decomposition $\vec n$ and $\vec n'$, and the variables $\xi$ and $\eta$, associated respectively to the extrinsic curvature and the area of a face.
These states are superposition of spinnetwork states, but they are peaked on a given geometry. 
In particular, it is possible to choose to peak them on a homogeneous and isotropic geometry. This corresponds to take a regular cellular decomposition. In particular, for the dipole, this corresponds to fix the normals and take all the spins being the same, so that we reduce to an unique holomorphic variable:
$ z_\ell = \xi_\ell +i\eta_\ell ~\to~ z=\alpha c +i\beta p$.
Geometry is determined by a couple $(c, p)$ in the past and $(c', p')$ in the future.
%

\section{Transition amplitude between an initial and a final universe}

Given these states, to which we have associated a cosmological interpretation, we can ask what is the transition amplitude between an initial and a final state. In LQG this transition amplitude is given as a sum over spinfoams $\sum_\sigma$, a product of ``face amplitude''  \cite{Bianchi:2010cr} and a product of vertex amplitudes $\prod_f d_f(\sigma) $, that reads \vspace{-11pt}
 $$~~~~~~~~~~~~~~
\bk W\psi=\sum_\sigma \ \prod_f d_f(\sigma)   \prod_v W_v(\sigma) ~.\vspace{-11pt}
$$
The form of this amplitude has been defined in  \cite{Livine:2007vk,Engle:2007wy,Freidel:2007py,Kaminski:2009fm}.
Here I focus on how to use it to calculate the transition amplitude associated to the transition from a 3-sphere to a 3-sphere. We use the dipole as initial and final boundary graph, and we choose the coherent states corresponding to a homogeneous isotropic geometry
.
We want to calculate this transition 
  at the first order. This corresponds to take a spinfoam with only one vertex.
%
%
%
For each vertex
\vspace{-3pt}
$$W(H_{\ell})= \int_{_{SO(4)^N}}\!\!\!\!\!\!\!\!\!\!\!\!d G_n \, ~
\prod_{\ell}     P_t(H_{\ell} \, ,
 \,
 G_{n_{source(\ell)}} G_{n_{target(\ell)}}^{-1} 
 )    ~~~~ \mbox{where}~~ G_n\!\in\!SO(4) ~~\forall\ \mbox{nodes}\ n
\vspace{-5pt}$$
and
\vspace{-3pt}
$$
P_t(H,G)\! =\! \! 
\sum_{j} {\scriptstyle (2j+1)}e^{\scriptscriptstyle-2t\hbar j(j+1)} \tr\!
\left[\! 
D^{\scriptscriptstyle(j)}\!(H)Y^\dagger D^{\scriptscriptstyle(j^{\!+}\!\!,j^{\!-}\!)}\!(G)Y
\! \right]   ~.
\vspace{-5pt}
$$
$D^{(j)}$ and $D^{(j^{\!+}\!,j^{\!-}\!)}$ are respectively the Wigner matrix of the spin-$j$ representation of $SU(2)$ and of the spin-$(j^{\!+}\!,j^{\!-}\!)$ representation of $SO(4)$. These matrices have different dimensions, and the map $Y$  glues them together.

Finally a further approximation is needed to obtain the semiclassical limit:  we consider the case of large spins $j$, equivalent to a large-distant expansion.
The calculation is now well-defined and can be performed \cite{Bianchi:2010uq}, and the result is this transition amplitude:
\bc \mbox{$
W(z,z')=C\ zz'e^{-\frac{z^2+(z')^2}\hbar}  ~.
$}\ec

Now, what does this transition amplitude mean physically?
 Our transition amplitude can be used directly to study the semiclassical behavior of the theory: in fact, semiclassical trajectories will be peaked on this amplitude (this can be seen numerically \cite{granada}). 
%
%
\\[1mm]
{\bf Effective Hamiltonian 
}~~~
We note that the transition amplitude is annihilated by the operator\\
$$
\hat H(z,\bar z
)\  W(z',z)= \f3{8\pi G}
 \left(\f{\hat z^2-\hat{\bar z}^2 -3t\hbar}{4\alpha\beta\gamma} \right)^{\!2} W(z',z)
=  \f3{8\pi G}
 \left(\f{4i\,\alpha\beta\,cp  -3t\hbar} {4\alpha\beta\gamma}\right)^{\!2} = 0~.
$$

In the large-distance approximation, where $p$ is larger than the Planck scale, one can disregard the term with $\hbar$, and we can use the fact that $p=a^2$ and $c$ corresponds to $\gamma\dot a$, thus  $\hat H \sim \f3{8\pi G}\ a^4\,\dot a^2$. We can divide by the volume in order to obtain the proper physical dimensions (since this is a constrain, it does not change the dynamics). We obtain $\hat H \sim \f3{8\pi G}\ a\,\dot a^2$
that is the usual Hamiltonian constraint for the FLRW dynamics.

This results shows that it is possible to derive the Friedmann equation from the full covariant loop gravity, and give a further proof of the robustness of the vertex used, since it gives a good semiclassical limit.

\section{Summary, comments and further developments}

To compute the transition amplitude we have used the expansions in the graph, the vertex and for large-distance.
We have clarified their role in selecting a geometry corresponding to some cosmology, and given two cosmological boundary states, we have seen that the corresponding transition amplitude leads to Friedmann dynamics in the semiclassical limit.

I want to stress the fact that the boundary states chosen are homogeneous and isotropic, but the Hilbert space in which they live, which is the Hilbert space of the dipole graph, contains more DOF than the isotropic and homogeneous ones. 
One of the advantage of this procedure, indeed, is to provide a model that take into account naturally the inhomogeneous and anisotropic DOF, and that can be use to describe the quantum geometry at the bounce, namely in a regime where we are interested to go beyond a single DOF. 
The main motivation in studying spinfoam cosmology is that it could provide the right setting to describe the quantum geometry at the bounce. In particular I want to understand if quantum fluctuations of the geometry  affect structure formation. In order to do this, one has to relax the assumptions of homogeneity and isotropy. In spinfoam cosmology is possible to choose how many degrees of freedom one wants to describe, by computing the dynamics on the appropriate graph. For this reason it would be interesting to investigate further the relation between the graph and the physical interpretation of the degrees of freedom involved. It could turn out to be very useful the to study this also in the  $U(N)$ formalism recently introduced     \cite{Freidel:2010zr,Borja:2010ly}.


In Spinfoam cosmology the next step to check the validity of the theory beyond the trivial flat solution is to include matter and/or a cosmological constant. One can include matter at an effective level, of course, but the coupling of matter with the quantum gravity is still an open task. On the other hand, there are already some developed ideas that can be followed to add a cosmological constant in spinfoam. 

Cosmology provides a privileged framework to study the relation between the canonical and the covariant LQG. There are several issues that can be explored. I would like to mention in particular the presence of ambiguities in the regularization of the Hamiltonian constraint. In LQC has been introduced an ``improved dynamics" \cite{Ashtekar:2006wn,Corichi:2008fu}, where the regularization is fixed by requiring a good semiclassical behavior. This choice can be made more robust if the resulting Hamiltonian will match with the one obtained in the covariant framework \cite{Bianchi:2010uq}.


\section*{References}
\providecommand{\newblock}{}

\end{document}